\documentclass[twocolumn,showpacs,amssymb,aps,nofootinbib,floatfix]{revtex4-1}
\usepackage{epsfig}
\usepackage{color}

\newcommand{\be}{\begin{eqnarray}}
\newcommand{\ee}{\end{eqnarray}}

\begin{document} \hbadness=10000
\topmargin -0.8cm\oddsidemargin = -0.7cm\evensidemargin = -0.7cm

\title{Heavy Quark Energy Loss in High Multiplicity Proton Proton Collisions at LHC}
\author{Sascha Vogel, Pol Bernard Gossiaux, Klaus Werner and J\"org Aichelin}
\address{SUBATECH,
Laboratoire de Physique Subatomique et des Technologies Associ\'ees \\
University of Nantes - IN2P3/CNRS - Ecole des Mines de Nantes \\
4 rue Alfred Kastler, F-44072 Nantes Cedex 03, France}

\begin{abstract}
One of the most promising probes to study deconfined matter created in high energy nuclear collisions is the energy loss of (heavy) quarks. It has been shown in experiments at the Relativistic Heavy Ion Collider (RHIC) that even charm and bottom quarks, despite their high mass, experience a remarkable medium suppression in the Quark Gluon Plasma.  \\
In this exploratory investigation we study the energy loss of heavy quarks in high multiplicity proton-proton collisions at LHC energies. Although the colliding systems are smaller than compared to those at RHIC (p+p vs. Au+Au) the higher energy might lead to multiplicities comparable to Cu+Cu collisions at RHIC. Recently the CMS collaboration has shown that these particles likely interact among each other. The interaction of charm quarks with this environment gives rise to a non-negligible suppression of high momentum heavy quarks in elementary collisions.

\end{abstract}

\maketitle


The start of the Large Hadron Collider (LHC) gives access to an unexplored area of high energy physics. While it has been commonly assumed that only high energy nucleus-nucleus collisions are able to produce a deconfined state of nuclear matter, commonly acknowledged as the Quark Gluon Plasma (QGP), the high energy and the resulting high multiplicities in proton-proton collisions at the LHC might allow for a QGP as well. The multiplicity of 7-14 TeV proton-proton collisions are comparable to Cu+Cu collisions at the Relativistic Heavy Ion Collider (RHIC) at Brookhaven National Laboratory (BNL).  Since one expects at least a partially deconfined state in Cu+Cu collisions at an center-of-mass energy of 200~AGeV \cite{Alver:2006wh, Lu:2008zzi, Abelev:2008jga,:2008sb,Lai:2009ai,Abelev:2010tr} it might be possible to reach a phase of deconfined matter in p+p collisions at the LHC as well. Recently the CMS collaboration indicated this possibility experimentally \cite{Khachatryan:2010gv}.    However, the proton-proton data always served as a baseline during the RHIC program and thus it might prove hard to find proper observables.

Heavy mesons would be an ideal probe to verify the existence of a QGP in proton proton collisions. If a deconfined plasma is created
and the heavy quarks pass this plasma, the same strong interaction between plasma constituents and heavy quarks
is expected which one has observed in RHIC heavy ion data. If no plasma is created and all hadrons are created as color-neutral particles in the vacuum no modification is expected. This argument originates from the fact that the interaction of the heavy mesons with the surrounding hadrons is much weaker compared to the strong interaction heavy quarks undergo in a deconfined state of matter. Thus a clear distinction for experimental signals is given. 

In this exploratory study we show that in very high multiplicity proton-proton collisions one can measure a small, however non-negligible  energy loss of heavy quarks. For our studies we use the model proposed in \cite{Gossiaux:2008jv,Gossiaux:2009hr,Gossiaux:2010yx} and adjust it to proton-proton collisions at LHC. The purpose of this letter is to demonstrate the effect of heavy quark energy loss in proton-proton collisions at LHC energies. It is organized as follows:
First, we will describe the models we use to calculate the underlying medium properties from the proton-proton collisions. We will then present the selection of high multiplicity proton-proton event within our theoretical framework. After that, we briefly review the elementary reactions that are the basis of our energy loss calculation. Finally we discuss the energy loss of heavy quarks in proton-proton collisions and discuss possible experimental observables. The letter ends with conclusions.

For our analysis, we split the medium description from the description of the elementary interaction of the heavy quarks with the medium (i.e. the light quarks and the gluons). Finally we combine both and calculate the spectra and a quantity similar to the nuclear suppression factor $R_{AA}$ in heavy ion collisions to specify the medium modification.\\
In order to simulate the proton-proton collisions and the medium that is created in those collisions we utilize the EPOS model including a hydro-dynamical phase, introduced in e.g. \cite{Werner:2010aa,Werner:2010ny,Werner:2010ss}. It has been thoroughly tested for proton-proton collision physics and is known to describe p+p collision data within reasonable bounds. 

\begin{figure}[htb]
\epsfig{width=6.9cm,clip=1,figure= 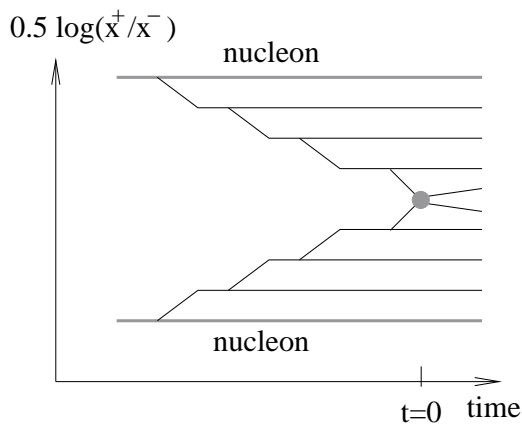}
\caption{(color online) Schematic view of a parton ladder.}
\label{partonladder}
\end{figure}

EPOS is a consistent quantum mechanical multiple scattering approach based on partons and strings \cite{Drescher:2000ha}, where cross sections and the particle production are calculated consistently, taking into account energy conservation in both cases (unlike other models where energy conservation is not considered for cross section calculations \cite{Hladik:2001zy}). A special feature is the explicit treatment of projectile and target remnants, leading to a very good description of baryon and antibaryon production as measured in proton-proton collisions at 158 GeV at CERN \cite{Liu:2003wja}. Nuclear effects related to CRONIN transverse momentum broadening, parton saturation, and screening have been introduced into EPOS \cite{Werner:2005jf}.
In heavy ion collisions (and more recently also in proton-proton collisions) collective behavior is taken into account \cite{Werner:2007bf}, in the following fashion: the initial scatterings, as described above, lead to the formation of strings, which break into segments, usually identified with hadrons. When it comes to heavy ion collisions, the procedure is modified: one considers the situation at an early proper time $\tau_0$, long before the hadrons are formed: one distinguishes between string segments in dense areas (more than some critical density $\rho_0$ segments per unit volume), from those in low density areas. The high density areas are referred to as core, the low density areas as corona \cite{Werner:2007bf}. It is important to note that initial conditions from EPOS are based on strings, providing a Òflux-tubeÓ like structure in case of individual events (a single flux tube in case of many overlaid events, for a schematic view of a fluxtube we refer to Fig. \ref{partonladder}). Based on the four-momenta of the string segments which constitute the core, we compute the energy density $\varepsilon(\tau_0,\vec{x}$) and the flow velocity $\vec{v}(\tau_0,\vec{x})$.
Having fixed the initial conditions, the system evolves according to the equations of ideal hydrodynamics. 

\begin{figure}[htb]
\epsfig{width=6.9cm,clip=1,figure= 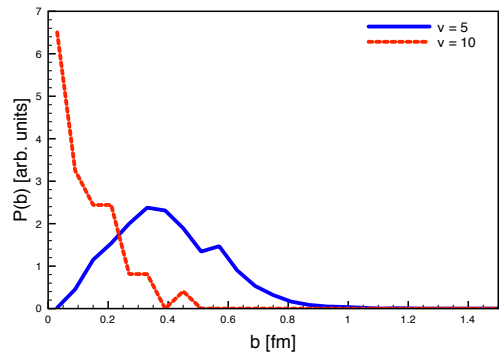}
\caption{(color online) Impact parameter distribution for two different choices of the number of elementary scatterings (fluxtubes).}
\label{impact_parameter}
\end{figure}

In order to identify high multiplicity proton-proton collisions we trigger on the number of elementary collisions (labeled $\nu$) within the proton-proton collision itself, i.e. the number of strings or color fluxtubes formed in the collisions. A fixed number of elementary collisions $\nu$ implies a distribution in multiplicity (as well as in impact parameter) which has a certain width. The distribution of impact parameters for two different choices of the number of binary collisions is shown in Fig. \ref{impact_parameter}. While this is not directly accessible within an experimental setup, an average multiplicity can be calculated and thus compared to experiment. For $\nu=5$ the average multiplicity of charged particles at mid-rapidity ${\mathrm d}N_{ch}/{\mathrm d}y|_{(y=0)}$ is 16.05, for collisions with  $\nu=10$ it equals 29.
We are calculating the energy loss for matter with an energy density larger than $1.5~ \mathrm {GeV / fm}^3$. Between $1.5~ \mathrm {GeV / fm}^3$ and $0.4~ \mathrm{GeV / fm}^3$ we are treating the matter as a so called mixed phase, with a reduced rate for reactions of heavy quarks with the medium. These values are in accordance with former hydrodynamical calculations applied to Au+Au reaction at RHIC. The impact of the hadronic phase has been checked and can be safely neglected. The interesting, possibly deconfined, phase is then roughly 2-3 fm in diameter (see e.g. \cite{Werner:2010ny}.) It has been cross-checked that for smaller values of the number of elementary interactions (i.e. smaller $\nu$, on the order of 1-2, which corresponds to multiplicities below 10) the medium modification of heavy quarks vanishes. 
A discussion on experimental observables can be found at the end of this letter.

\begin{figure}[htb]
\epsfig{width=6.9cm,clip=1,figure= 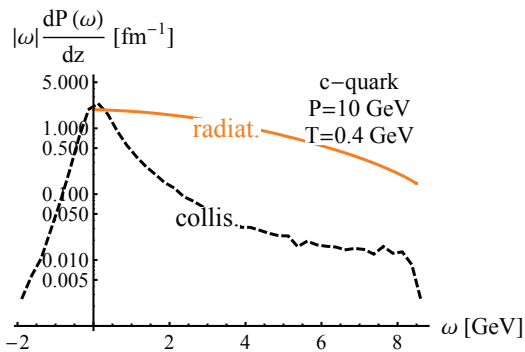}
\caption{(color online) Probability of collisional (dashed line) and radiative (solid line) energy loss per unit length as a function of energy loss $\omega$.}
\label{eloss}
\end{figure}

The energy loss model we use for our studies is MC@sHQ, for details we refer to \cite{Gossiaux:2008jv,Gossiaux:2009hr,Gossiaux:2010yx}. For brevity we do only repeat the key features of the model. Except for adjusting aspects of the technical implementation the physical picture remains the same in p+p collisions as in Au+Au collisions.\\
The two major improvements in comparison to conventional pQCD inspired models are the inclusion of a running coupling constant and a proper matching of the infrared regulator of the gluon propagator to HTL calculations. These changes allowed to describe data in heavy ion collisions at RHIC within reasonable K-factors. For a recent comparison to experimental data, we refer to \cite{Gossiaux:2010yx}. Additionally to collisional energy loss a radiative component to the energy loss is implemented. This is calculated as incoherent gluon bremsstrahlung pushing to its limit the idea that the gluon formation time is strongly reduced when the gluon is radiated off heavy quarks. For more information we refer to \cite{Gossiaux:2010yx}.  The probability of collisional or radiative energy lost per unit length for a 10 GeV c-quark is shown in Fig. \ref{eloss} as a function of energy. 
Collisional energy loss (dashed line) mostly consists of a large number of collisions in which the incoming heavy quark loses a small part of its energy, hence the pronounced peak at around zero in Fig. \ref{eloss}. For radiative losses (solid line) the probability to lose a larger fraction of the heavy quark energy in a single process is more important.

To combine the description of the medium and the energy loss, we generate an averaged medium, i.e. an energy density profile with EPOS. Within this medium heavy quarks are generated and then propagated according to the Boltzmann equation. We then compare two different scenarios. First, the heavy quarks lose energy according to the description in \cite{Gossiaux:2010yx}. Alternatively the heavy quarks move freely in the medium. The ratio of both final spectra -- that is, the quenching --  will then provide a measure for the lost energy.\\

\begin{figure}[htb]
\epsfig{width=6.9cm,clip=1,figure= 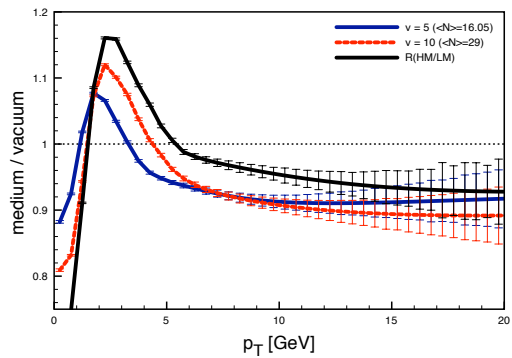}
\caption{(color online) The $R_{AA}$ equivalent in proton-proton collisions for different choices of the number of elementary scatterings (fluxtubes). For $\nu=5$ the average multiplicity of charged particles at mid-rapidity ${\mathrm d}N_{ch}/{\mathrm d}y|_{(y=0)}$ is 16.05, for collisions with  $\nu=10$ it equals 29. The curve labeled $R_{HM/LM}$ depicts a high multiplicity spectrum ($\nu$ = 10, ${\mathrm d}N_{ch}/{\mathrm d}y|_{(y=0)}$=29) divided by a low multiplicity spectrum ($\nu$=1, ${\mathrm d}N_{ch}/{\mathrm d}y|_{(y=0)}=3.62$). Error bars shown are statistical.}
\label{RAA}
\end{figure}

In Fig. \ref{RAA} we show the quenching (the equivalent of a nuclear suppression factor $R_{AA}$ for proton-proton collisions) for charm quarks. Theoretically this is well-defined, since one can calculate the medium-influenced spectra as well as the vacuum spectra by neglecting reinteraction of the heavy quark with the medium. The ratio of both then gives the desired quantity. Additionally the below defined observable $R_{HM/LM}$ is shown, which shows a very similar suppression. One should note that the suppression for $\nu = 5$ and $\nu = 10$ seem very similar. This originates from the fact that the heavy quark undergoes few interactions due to the generally small interaction zone. Once the energy density reaches a certain threshold the suppression sets in and the particles with high transverse momentum escape the high energy density zone rapidly. Thus the low $p_T$ part of the spectrum is effected more, which is in line with the observation in Fig. \ref{RAA}. 

One observes a small, however non-negligible energy loss of charm quarks at high transverse momentum manifesting itself as a 10\% suppression of the spectra at large $p_T$ ($p_T>$ 7GeV). This prediction should however be considered
with a grain of salt: it is known -- for a recent review, see
\cite{Peigne:2009} -- that the heavy quarks produced inside a hot medium undergo a transient 
regime before losing energy proportional to the path-length, an effect not implemented 
in our simulations. Also additional initial state effects might have shift the data in the opposite direction and more systematic studies are possible once LHC data arrives.\\

When trying to define this observable for experimental applications, we are confronted with the complication that the reference measurement uses the same colliding systems as the medium measurement. Thus we need a quantity which refers to experimental observables. 
We suggest a new ratio, which can be measured experimentally.
It requires knowledge of multiplicity of the proton-proton collision, and a clear cut between low multiplicity and high multiplicity collisions. Then one can construct an observable similar to $R_{CP}$ in heavy ion collisions \cite{Adler:2003kg}.
Measuring the spectra in high multiplicity p+p collisions and dividing them by the spectra obtained from low multiplicity collisions one constructs 

\begin{equation}
R_{HM/LM} = \frac{\frac{dN/dp_{T,HM}}{N_{HM}}}{\frac{dN/dp_{T,LM}} {N_{LM}}} .
\end{equation}

Here, $dN/dp_{T,HM}$ denotes the HQ spectrum obtained from high multiplicity proton-proton collisions, $dN/dp_{T,LM}$ is the spectrum as obtained from low multiplicity proton-proton collisions. $N_{HM}$ and $N_{LM}$ denote the multiplicities in high multiplicity or low multiplicity proton-proton collisions respectively. In our analysis we assume  the initial HQ spectrum scales linearly with the number of binary reactions.  The advantage of this observable is that it is "self-normalizing", due to the scaling with the respective multiplicities, which serve as a normalization of the spectra. In principle one should aim for the bins with the highest and lowest multiplicity, for our analysis we have chosen two values ${\mathrm d}N_{ch}/{\mathrm d}y|_{(y=0)}$ = 29 ($\nu = 10$) and 16.05 ($\nu = 5$) in the high multiplicity range to compare possible effects. For the low multiplicity bin we selected the one with ${\mathrm d}N_{ch}/{\mathrm d}y|_{(y=0)}$ = 3.62 ($\nu = 1$), however experimental statistics might be enough to aim for even higher multiplicities. Apart from experimental difficulties we predict a quenching of roughly 10\% at $p_T > 10~\mathrm{GeV}$ in high multiplicity proton-proton collisions as indicated in Fig. \ref{RAA}. \\

Instead of a low multiplicity sample one could choose minimum bias data as well, since it is dominated by collisions producing few number of particles. Although this might be more accessible to experiments, we suggest that a clear cut between low and high multiplicity collisions enhances the viability of this observable.

Another possible scaling is a scaling with $z = N_{ch} / \left<N_{ch}\right>$, where $N_{ch}$ denotes the number of charged particles in a certain centrality class and $\left<N_{ch}\right>$ the mean value from minimum bias (MB) data. A possible experimental observable would be
\begin{equation}
R(z)=  \left.\frac{dN/dp_T (N_{ch})}{dN/dp_T (MB)}   \right|_{p_T>10 GeV} \times \frac{N_{MB}}{N_{ch}} , 
\end{equation}
which can be evaluated in different multiplicity classes and thus different values of z.
The scaling with z might give some insight into the centrality dependence of the energy loss of heavy quarks. 

\begin{figure}[t]
\epsfig{width=6.9cm,clip=1,figure= 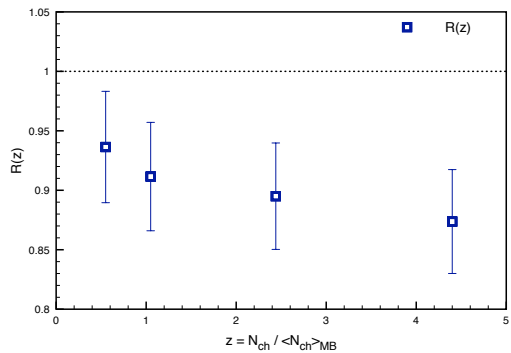}
\caption{(color online) High transverse momentum suppression of charm quarks as a function of $z = N_{ch} / \left<N_{ch}\right>$. Shown is the medium calculation divided by the vacuum calculation, averaged above 10 GeV. Error bars shown are statistical. }
\label{rz}
\end{figure}

Shown in Fig. \ref{rz} is a calculation of R(z) within our approach. One observes a decrease of roughly 10-15\% with increasing centrality/multiplicity of the collision, which is in line with the expectation from the previous arguments. 

\section*{Conclusions}
In our exploratory study we have shown that if in high multiplicity proton-proton collisions a deconfined phase is produced, heavy quarks will be a suitable probe for its experimental analysis. If fast heavy quarks pass a deconfined medium they lose energy due to the interaction with the medium and thus are quenched. This quenching reaches about 10\% at large transverse momentum. By dividing the heavy quark spectra for high and low multiplicity p+p collisions (scaled accordingly) this effect can be experimentally observed at LHC experiments. An insightful measurement would be the energy loss as a function of multiplicity, e.g. as a function of the variable $z = N_{ch}/\left<N_{ch}\right>$. By this measurement a clear onset of medium formation in proton-proton collisions could be observed.

\section*{Acknowledgements}
Computational resources have been provided by Subatech and Centre de Calcul de l'IN2P3.

\end{document}